\documentclass[twocolumn,aps,prc,showpacs,floatfix]{revtex4}
\usepackage{graphicx}
\usepackage{dcolumn}
\usepackage{bm}
\usepackage{CJK}

\begin{document}
\begin{CJK*}{GBK}{song}

\preprint{}
\title{Probing nuclear symmetry energy with the sub-threshold pion production}
\author{Zhang Fang}
\affiliation{School of Nuclear Science and Technology, Lanzhou
University, Lanzhou 730000, China}
\author{Liu Yang, Yong Gao-Chan and Zuo Wei }\affiliation{Institute of
Modern Physics, Chinese Academy of Sciences, Lanzhou 730000,
China}

%\date{\today}

\begin{abstract}
Within the framework of semiclassical Boltzmann-Uehling-Uhlenbeck
(BUU) transport model, we investigated the effects of symmetry
energy on the sub-threshold pion using the isospin MDI interaction
with the stiff and soft symmetry energies in the central collision
of $^{48}$Ca + $^{48}$Ca at the incident beam energies of 100,
150, 200, 250 and 300 MeV/nucleon, respectively. We find that the
ratio of $\pi^{-}/\pi^{+}$ of sub-threshold charged pion
production is greatly sensitive to the symmetry energy,
particularly around 100 MeV/nucleon energies. Large sensitivity of
sub-threshold charged pion production to nuclear symmetry energy
may reduce uncertainties of probing nuclear symmetry energy via
heavy-ion collision.
\end{abstract}

 \maketitle

%\section{Introduction}
The density dependence of the nuclear symmetry energy is not only
important for nuclear physics, but also crucial for many
astrophysical processes, such as the structure of neutron stars
and the dynamical evolution of proto-neutron stars \cite{M97}.
Though considerable progress has been made recently in determining
the density dependence of the nuclear symmetry energy around the
normal nuclear matter density from studying the isospin diffusion
in heavy-ion reactions at intermediate energies
\cite{tsang04,chen05,li05}, much more work is still needed to
probe the high-density behavior of the nuclear symmetry energy.
Fortunately, heavy-ion reactions, especially those induced by
radioactive beams, provide a unique opportunity to constrain the
EOS of asymmetric nuclear matter, and a number of such observables
have been already identified in heavy-ion collisions induced by
neutron-rich nuclei, such as the free neutron/proton ratio
\cite{li03}, the isospin fractionation \cite{h00,w01}, the
neutron-proton transverse differential flow \cite{li00}, the
neutron-proton correlation function \cite{chen03}, t/3He
\cite{y05}, the isospin diffusion \cite{l03}, the proton
differential elliptic flow \cite{li01}. Currently, to pin down the
symmetry energy, the National Superconducting Cyclotron Laboratory
(NSCL) at Michigan State University, the Geselschaft fuer
Schwerionenforschung (GSI) at Darmstadt, the Rikagaku Kenkyusho
(RIKEN, The Institute of Physical and Chemical Research) of Japan,
as well as the Cooler Storage Ring (CSR) in Lanzhou, are planning
to do related experiments. Some of related information can be find
via Ref.~\cite{tsang2012}.

Recently, pion production in heavy-ion collisions has attracted
much attention in the nuclear physics community
\cite{xiao09,di10,yong06,yong2011}. One important reason for this
is that pion production is connected with the high-density
behavior of nuclear symmetry energy, especially around pion
production threshold. The latter is crucial for understanding many
interesting issues in both nuclear physics and astrophysics. And
several hadronic transport models have quantitatively shown that
$\pi^{-}/\pi^{+}$ ratio is indeed sensitive to the symmetry
energy, especially around pion production threshold \cite{yong06}
and above pion production threshold \cite{feng}. In the previous
studies, pion production mainly from energetic heavy-ion
collisions, in which pions are mainly produced above their
threshold energy. What is the case of effects of nuclear symmetry
energy on charged pion ratio via sub-threshold production? To
answer this question, we studies the effects of nuclear symmetry
energy on sub-threshold pion production. We select
$^{48}$Ca+$^{48}$Ca as the reaction system due to it large
asymmetry. Based on the isospin-dependent Boltzmann-Uehling-
Uhlenbeck (IBUU) transport model, we studied the effects of the
symmetry energy in the central reaction $^{48}$Ca+$^{48}$Ca and
find that the ratio of $\pi^{-}/\pi^{+}$ of sub-threshold charged
pion production, compared with above threshold case, is
particularly sensitive to the symmetry energy.

%\section{THE IBUU04 TRANSPORT MODEL}
Our present work is based on the semi-classical transport model
IBUU04, in which nucleons, $\Delta $ and $N^{\ast }$ resonances as
well as pions and their isospin-dependent dynamics are included.
We use the isospin-dependent in-medium nucleon-nucleon ($NN$)
elastic cross sections from the scaling model according to nucleon
effective masses. For the inelastic cross sections we use the
experimental data from free space $NN$ collisions since at higher
incident beam energies the $NN$ cross sections have no evident
effects on the slope of neutron-proton differential flow. The
total and differential cross sections for all other particles are
taken either from experimental data or obtained by using the
detailed balance formula. The isospin dependent phase-space
distribution functions of the particles involved are solved by
using the test-particle method numerically. The isospin-dependence
of Pauli blockings for fermions is also considered. The
isospin-dependence of Pauli blockings for fermions is also
considered. More details can be found in Refs.
 \cite{li05,li03,li04,das03,l04,yong06}. In the present studies, the
momentum-dependent single nucleon potential (MDI) adopted here
is:%
\begin{eqnarray}
U(\rho ,\delta ,\mathbf{p},\tau ) &=&A_{u}(x)\frac{\rho _{\tau ^{\prime }}}{%
\rho _{0}}+A_{l}(x)\frac{\rho _{\tau }}{\rho _{0}}  \nonumber \\
&&+B(\frac{\rho }{\rho _{0}})^{\sigma }(1-x\delta ^{2})-8x\tau \frac{B}{%
\sigma +1}\frac{\rho ^{\sigma -1}}{\rho _{0}^{\sigma }}\delta \rho _{\tau
^{\prime }}  \nonumber \\
&&+\frac{2C_{\tau ,\tau }}{\rho _{0}}\int d^{3}\mathbf{p}^{\prime }\frac{%
f_{\tau }(\mathbf{r},\mathbf{p}^{\prime })}{1+(\mathbf{p}-\mathbf{p}^{\prime
})^{2}/\Lambda ^{2}}  \nonumber \\
&&+\frac{2C_{\tau ,\tau ^{\prime }}}{\rho _{0}}\int d^{3}\mathbf{p}^{\prime }%
\frac{f_{\tau ^{\prime }}(\mathbf{r},\mathbf{p}^{\prime })}{1+(\mathbf{p}-%
\mathbf{p}^{\prime })^{2}/\Lambda ^{2}}.  \label{potential}
\end{eqnarray}%
In the above equation, $\delta =(\rho _{n}-\rho _{p})/(\rho
_{n}+\rho _{p})$ is the isospin asymmetry parameter, $\rho =\rho
_{n}+\rho _{p}$ is the baryon density and $\rho _{n},\rho _{p}$
are the neutron and proton densities, respectively. $\tau
=1/2(-1/2)$ for neutron (proton) and $\tau \neq \tau ^{\prime }$,
$\sigma =4/3$, $f_{\tau }(\mathbf{r},\mathbf{p})$ is the
phase-space distribution function at coordinate $\mathbf{r}$ and
momentum $\mathbf{p}$. The parameters $A_{u}(x),A_{l}(x),B,C_{\tau
,\tau }$, $C_{\tau ,\tau ^{\prime }}$ and $\Lambda $ were set by
reproducing the momentum-dependent potential $U(\rho ,\delta
,\mathbf{p},\tau )$ predicted by the Gogny Hartree-Fock and/or the
Brueckner-Hartree-Fock calculations. The momentum-dependence of
the symmetry potential steams from the different interaction
strength parameters $C_{\tau,\tau'}$ and $C_{\tau,\tau}$ for a
nucleon of isospin $\tau$ interacting, respectively, with unlike
and like nucleons in the background fields, more specifically,
$C_{unlike}=-103.4$ $MeV$ while $C_{like}=-11.7$ $MeV$. The
parameters $A_{u}(x)$and $A_{l}(x)$ depend on the $x$ parameter
according to $Au(x)=-95.98-x\frac{2B}{\sigma+1}$ and $A_{l}(x) =
-120.57+x\frac{2B}{\sigma+1}$. The saturation properties of
symmetric nuclear matter and the symmetry
energy of about $32$ MeV at normal nuclear matter density $\rho _{0}=0.16$ fm%
$^{-3}$. The incompressibility of symmetric nuclear matter at
normal density is set to be $211$ MeV. According to essentially
all microscopic model calculations, the EOS for isospin asymmetric
nuclear matter can be expressed as
\begin{equation}
E(\rho ,\delta )=E(\rho ,0)+E_{\text{sym}}(\rho )\delta ^{2}+\mathcal{O}%
(\delta ^{4}),
\end{equation}%
where $E(\rho ,0)$ and $E_{\text{sym}}(\rho )$ are the energy per
nucleon of symmetric nuclear matter and nuclear symmetry energy,
respectively. For a given value $x$, with the single particle
potential $U(\rho ,\delta,\mathbf{p},\tau )$, one can readily calculate the symmetry energy $E_{\text{sym}%
}(\rho )$ as a function of density.

The main reaction channels related to pion production and
absorption are
\begin{eqnarray}
&& NN \longrightarrow NN \nonumber\\
&& NR \longrightarrow NR \nonumber\\
&& NN \longleftrightarrow NR \nonumber\\
&& R \longleftrightarrow N\pi,
\end{eqnarray}
where $R$ denotes $\Delta $ or $N^{\ast }$ resonances. In the
present work, we use the isospin-dependent in-medium reduced $NN$
elastic scattering cross section from the scaling model according
to nucleon effective mass \cite{yong06} to study the effect of
symmetry energy on pion production. Assuming in-medium $NN$
scattering transition matrix is the same as that in vacuum, the
elastic $NN$ scattering cross section in medium $\sigma
_{NN}^{medium}$ is reduced compared with their free-space value
$\sigma _{NN}^{free}$ by a factor of \cite{epja}
\begin{eqnarray}
R_{medium}(\rho,\delta,\textbf{p})&\equiv& \sigma
_{NN_{elastic}}^{medium}/\sigma
_{NN_{elastic}}^{free}\nonumber\\
&=&(\mu _{NN}^{\ast }/\mu _{NN})^{2}.
\end{eqnarray}
where $\mu _{NN}$ and $\mu _{NN}^{\ast }$ are the reduced masses
of the colliding nucleon pair in free space and medium,
respectively. For in-medium $NN$ inelastic scattering cross
section, even assuming in-medium $NN \rightarrow NR$ scattering
transition matrix is the same as that in vacuum, the density of
final states $D_{f}^{'}$ of $NR$ is very hard to calculate due to
the fact that the resonance's potential in matter is presently
unknown. The in-medium $NN$ inelastic scattering cross section is
thus quite controversial. Because the purpose of present work is
just study the effect of symmetry energy on pion production and
charged pion ratio, to simplify the question, for the $NN$
inelastic scattering cross section we use the free $NN$ inelastic
scattering cross section. The effective mass of nucleon in isospin
asymmetric nuclear matter is
\begin{equation}
\frac{m_{\tau }^{\ast }}{m_{\tau }}=\left\{ 1+\frac{m_{\tau }}{p}\frac{%
dU_{\tau }}{dp}\right\}^{-1}.
\end{equation}
From the definition and Eq.~(\ref{potential}), we can see that the
effective mass depends not only on density and asymmetry of medium
but also the momentum of nucleon.

%\section{Results and discussions}

\begin{figure}[th]
\begin{center}
\includegraphics[width=0.5\textwidth]{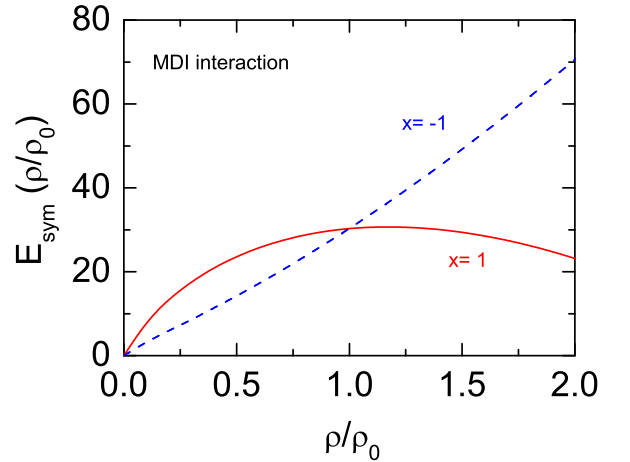}
\end{center}
\caption{{\protect\small (Color online) Density dependence of
nuclear symmetry energy with parameters $x= 1, -1$,
respectively.}} \label{Esym}
\end{figure}
\begin{figure}[th]
\begin{center}
\includegraphics[width=0.5\textwidth]{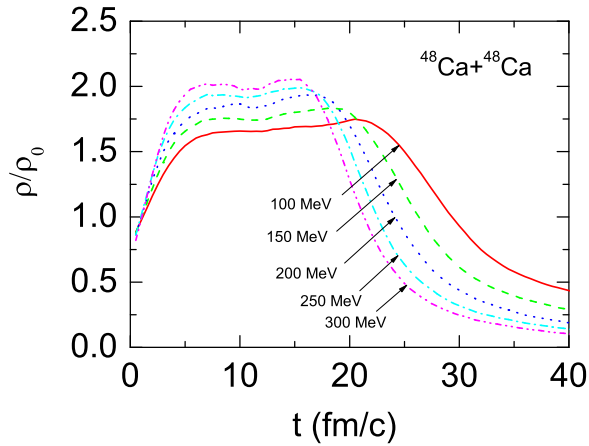}
\end{center}
\caption{{\protect\small (Color online) Maximal baryon density
reached in the central reaction $^{48}$Ca+$^{48}$Ca with different
incident beam energies.}} \label{den}
\end{figure}
Fig.\ \ref{Esym} shows the density dependence of nuclear symmetry
energy with parameter $x=1, -1$, respectively. As discussed in the
previous part, the single particle used has an $x$ parameter,
different specific $x$ parameter denotes different density
dependent symmetry energy. For the central reaction
$^{48}$Ca+$^{48}$Ca, the maximal density reached is about 1.5
$\sim$ 2 times saturation density as shown in Fig.\ \ref{den}.
From Fig.\ \ref{Esym}, we can also see that the low density
behaviors of nuclear symmetry energy separate from each other with
different $x$ parameters. At the saturation point there is a cross
and then they separate from each other again. At lower densities,
the value of symmetry energy of $x=0$ is lower than that of $x=1$.
But at high densities, the value of symmetry energy of $x=0$ is
higher than that of $x=1$. From Fig.\ \ref{den}, we can see that
the maximal baryon density reached in the central reaction
$^{48}$Ca+$^{48}$Ca increases with the incident beam energy. At
100 MeV/nucleon, the maximal baryon density reaches about 1.5
times saturation density and at 300 MeV/nucleon, the maximal
baryon density reaches about 2 times saturation density. Therefore
the ratio of $\pi^{-}/\pi^{+}$ of sub-threshold charged pion
production still mainly reflects the high density behavior of
nuclear symmetry energy.

\begin{figure}[th]
\begin{center}
\includegraphics[width=0.5\textwidth]{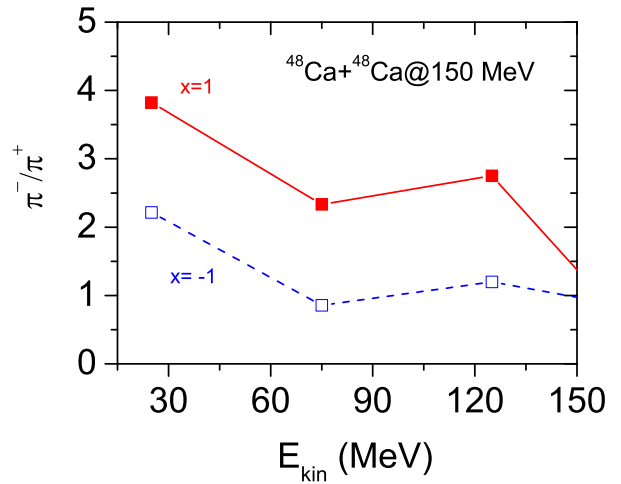}
\end{center}
\caption{{\protect\small (Color online) Kinetic energy
distribution of the $\pi^{-}/\pi^{+}$ ratio using the MDI
interaction with $x=1$ and -1 for the central collision of
$^{48}$Ca + $^{48}$Ca at the incident beam energies of 150
MeV/nucleon.}} \label{k150}
\end{figure}
\begin{figure}[th]
\begin{center}
\includegraphics[width=0.5\textwidth]{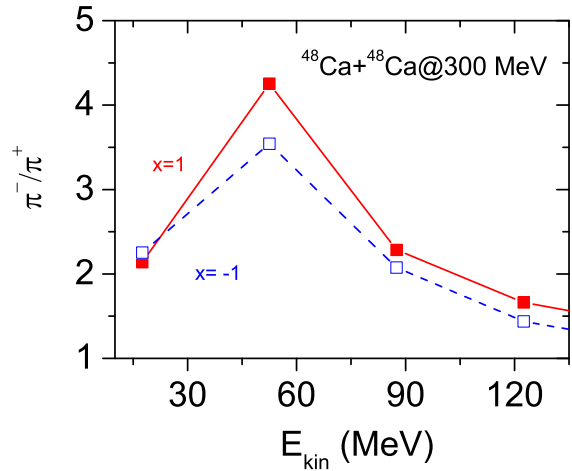}
\end{center}
\caption{{\protect\small (Color online) Kinetic energy
distribution of the $\pi^{-}/\pi^{+}$ ratio using the MDI
interaction with $x=1$ and -1 for the central collision of
$^{48}$Ca + $^{48}$Ca at the incident beam energies of 300
MeV/nucleon.}} \label{k300}
\end{figure}
Fig.\ \ref{k150} shows the kinetic energy distribution of the
$\pi^{-}/\pi^{+}$ ratio using the MDI interaction with $x=1$ and
-1 for the central collision of $^{48}$Ca + $^{48}$Ca at the
incident beam energies of 150 MeV/nucleon. It is seen that at pion
kinetic energies of 30 $\sim$ 120 MeV, the $\pi^{-}/\pi^{+}$ ratio
is very sensitive to the symmetry energy. The soft symmetry energy
(x= 1) gives larger $\pi^{-}/\pi^{+}$ ratio whereas the stiff
symmetry energy ($x=-1$) gives smaller $\pi^{-}/\pi^{+}$ ratio.
This is consistent with the previous studies
\cite{xiao09,yong06,yong2011} for the neutron-rich reactions. And
we can see that the sensitivity of $\pi^{-}/\pi^{+}$ ratio to the
high density behavior of symmetry energy is quite large, about
80\%, is quite sensitive to cases of pion production at higher
incident beam energies \cite{yong05}. The sub-threshold charged
pion production is thus a very sensitive probe of the symmetry
energy. Fig.\ \ref{k300} is the case at 300 MeV/nucleon incident
beam energy. Again we clearly see that at pion kinetic energies of
30 $\sim$ 120 MeV, the $\pi^{-}/\pi^{+}$ ratio is sensitive to the
symmetry energy. The soft symmetry energy ($x=1$) gives larger
$\pi^{-}/\pi^{+}$ ratio whereas the stiff symmetry energy ($x=-1$)
gives smaller $\pi^{-}/\pi^{+}$ ratio. Sensitivity of charged pion
ratio to the symmetry energy at higher incident beam energy 300
MeV/nucleon is smaller than the case at lower incident beam energy
150 MeV/nucleon. In the heavy-ion collisions at intermediate
energies, generally speaking, mean-field effect and collision
effect compete each other. The effect of mean-field increases with
the decreasement of incident beam energy. Thus at lower incident
beam energies effects of the symmetry energy (which is the
isovector part of nuclear mean-field potential) are larger than
that with higher incident beam energies. From Fig.\ \ref{k300}, we
can also see that there is an evident Coulomb peak \cite{yong06},
and around the Coulomb peak $\pi^{-}/\pi^{+}$ ratio is more
sensitive to the symmetry energy. At lower incident beam energy
150 MeV/nucleon, however, there is no Coulomb peak at at pion
kinetic energies of 30 $\sim$ 120 MeV.

\begin{figure}[th]
\begin{center}
\includegraphics[width=0.5\textwidth]{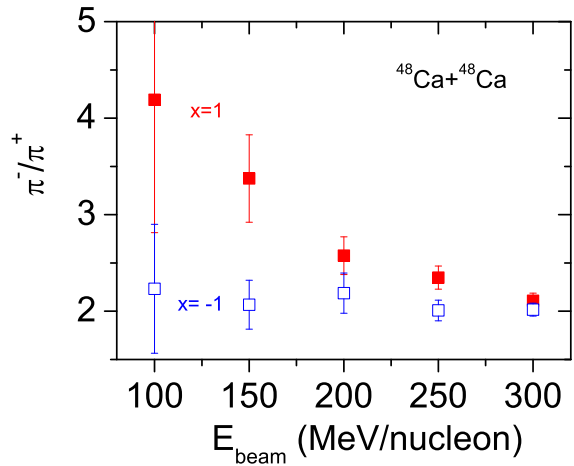}
\end{center}
\caption{{\protect\small (Color online) Excitation function of the
$\pi^{-}/\pi^{+}$ ratio using the MDI interaction with $x=1$ and
-1 for the central collision of $^{48}$Ca + $^{48}$Ca at the
incident beam energies of 100, 150, 200, 250 and 300 MeV/nucleon,
respectively.}} \label{rpion}
\end{figure}
Fig.\ \ref{rpion} shows the excitation function of the
$\pi^{-}/\pi^{+}$ ratio using the MDI interaction with $x=1$ and
-1 for the central collision of $^{48}$Ca + $^{48}$Ca at the
incident beam energies of 100, 150, 200, 250 and 300 MeV/nucleon,
respectively. From Fig.\ \ref{rpion} we can clearly see that
sensitivity of charged pion ratio $\pi^{-}/\pi^{+}$ reaches the
maximum at the lower incident beam energy 100 MeV/nucleon, about
100\% sensitive to the symmetry energy. The trend of the
sensitivity of the charged pion ratio $\pi^{-}/\pi^{+}$ to
symmetry energy decreases with the incident beam energy is
consistent with the previous studies \cite{xiao09}. Note here that
at the incident beam energy region studied here, the soft symmetry
energy always corresponds large value of $\pi^{-}/\pi^{+}$ ratio
and the stiff symmetry energy corresponds relative small value of
$\pi^{-}/\pi^{+}$ ratio \cite{eos2011}. However, in Ref.
\cite{feng}, Feng et al. systematically investigated the pion
production in heavy-ion collisions in the region of below 1
\emph{A}GeV energies by using the ImIQMD model. They found that
the excitation functions of the $\pi^{-}/\pi^{+}$ ratio increases
with the stiffness of the symmetry energy. This is inconsistent
with our studies \emph{qualitatively} and the reasons are needed
to be clarified.

%\section{Summary}

In summary, based on the semiclassical Boltzmann-Uehling-Uhlenbeck
(BUU) transport model, we studied the effects of symmetry energy
on the sub-threshold pion production in the central reaction
$^{48}$Ca+$^{48}$Ca. We find that the ratio of $\pi^{-}/\pi^{+}$
of sub-threshold charged pion production is particularly sensitive
to the symmetry energy with decrease of incident beam energy of
heavy-ion collision. The highly sensitive charged sub-threshold
pion ratio to the symmetry energy may reduce the uncertainties of
probing nuclear symmetry energy via heavy-ion collision.

%\section*{Acknowledgments}

The work is supported by the National Natural Science Foundation
of China (10875151, 10947109, 11175219, 10740420550), the
Knowledge Innovation Project (KJCX2-EW-N01) of Chinese Academy of
Sciences, the Major State Basic Research Developing Program of
China under No.2007CB815004, the CAS/SAFEA International
Partnership Program for Creative Research Teams (CXTD-J2005-1),
the Fundamental Research Fund for Physics and Mathematic of
Lanzhou University LZULL200908, the Fundamental Research Funds for
the Central Universities lzujbky-2010-160.

\end{CJK*}

\end{document}